\documentclass[10pt,aps,prb,twocolumn,superscriptaddress]{revtex4}

\usepackage{epsfig}
\usepackage{dcolumn}
\usepackage{bm}
\usepackage{amsmath}
\usepackage{amsfonts}
\usepackage{latexsym}
\usepackage{amssymb}
\usepackage{color}
\usepackage{hyperref}
\usepackage[normalem]{ulem}

\newcommand{\g}[1]{{\bf #1}}
\newcommand{\us}{USb$_2$}
\newcommand{\ua}{UAs$_2$}

\newcommand{\ub}{UBi$_2$}
\newcommand{\ug}{UGe$_2$}
\newcommand{\ux}{UX$_2$}

\newcommand{\be}{\begin{equation}}
\newcommand{\ee}{\end{equation}}
\newcommand{\bea}{\begin{eqnarray}}
\newcommand{\eea}{\end{eqnarray}}
\newcommand{\ba}{\begin{eqnarray*}}
\newcommand{\ea}{\end{eqnarray*}}
\newcommand{\dagga}{{\phantom{\dagger}}}

\begin{document}

\title{Microscopic mechanism for the unusual antiferromagnetic order\\ and the pressure-induced transition to ferromagnetism in \us}

\author{Marcin M. Wysoki\'nski}
\email{mwysoki@sissa.it}

\affiliation{International School for Advanced Studies (SISSA)$,$ via Bonomea 265$,$ IT-34136$,$ Trieste$,$ Italy}
\affiliation{Marian Smoluchowski Institute of Physics$,$ Jagiellonian
University$,$ 
ulica prof. S. \L ojasiewicza 11$,$ PL-30348 Krak\'ow$,$ Poland}
 \affiliation{International Research Centre MagTop at Institute of
Physics, Polish Academy of Sciences, Aleja Lotnik\'ow 32/46,
PL-02668 Warszawa, Poland}

\date{\today}

\begin{abstract}
Uranium dipnictide USb$_2$ reflects enigmatic properties posing a substantial challenge for a microscopic modeling. Among others, it develops a nonstandard antiferromagnetic order  of a $\uparrow\downarrow\downarrow\uparrow$-type along [001] crystallographic direction, and  under pressure it undergoes transition to the ferromagnetic phase. Here we propose a minimal low-energy model of USb$_2$  which, as we demonstrate at the mean-field level, accommodates physical mechanism for mentioned observations.
Relying on the obtained results we also comment on the features of magnetism observed in other U-based compounds: UAs$_2$, UBi$_2$, UAsSe, URh$_x$Ir$_{1-x}$Ge and UGe$_2$.
\end{abstract}

\maketitle

\section{Introduction} 
Compounds with active $f$-electron degrees of freedom host a wealth of diverse unconventional phases of matter.
These  encompass unconventional superconductivity \cite{Pfleiderer2009,Aoki2012}, magnetism \cite{Brando2016}, and many-body, also possibly topologically nontrivial \cite{Coleman2010,Coleman2017,Wysokinski2016R}, insulating and metallic states \cite{Riseborough2000,Stewart1984}. However, small energy scales associated with specific features of the electronic structure near Fermi level of $f$-electron materials  substantially limit the effectiveness of many experimental techniques probing the character of emergent orders. 

In this light, among various $f$-electron compounds, \us\  stands out with the relatively accurately studied electronic structure. 
There exists a wealth of observations  made for this material with a broad spectrum of experimental techniques including: photo-emission \cite{Guziewicz2004,Durakiewicz2008,Beaux2011, Durakiewicz2013usb,Xie2016}, M\"ossbauer \cite{Tsutsui2004} and hard X-ray \cite{Beaux2011} spectrocopies as well as  neutron diffraction \cite{Leciejewicz1967}, de Haas - van Alphen oscillations \cite{Aoki1999,Aoki2000}, nuclear magnetic resonance \cite{Kato2004,Baek2010} and very recent magnetization, magnetostriction and magnetotransport measurements under extreme conditions of high fields \cite{Jeffries2017} and strong pressures \cite{Jeffries2016}.

\begin{figure}[b]
  \begin{center}  
    \includegraphics[width= 0.35   \textwidth]{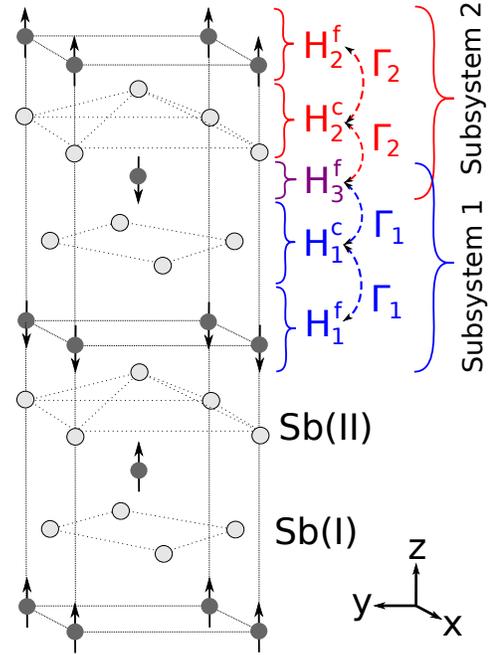}
  \end{center} 
 \caption{Schematic picture of the magnetic unit cell of \us\ with marked orientations of spins at uranium atoms (dark gray color). Inequivalent ligand sites (light gray color) are marked as Sb(I) and Sb(II) \cite{Baek2010,Beaux2011}. 
 The curly brackets underline assignments of various terms in Eq. \eqref{H} for a description of a \us\ film (parallel to $xy$-plane) of thickness corresponding to the height of the elementary cell.    
}\vspace{-0.1	cm}\label{fig1} 
\end{figure}

In spite of thorough experimental examination of \us, faithful theoretical understanding of some of its characteristics is up to the present day offered only by {\it ab-inito} studies \cite{Eriksson2006}. These, however, allow only for a limited insight into physical mechanisms responsible for the specific behavior of a material. In particular, the origin of the nonstandard $\uparrow\downarrow\downarrow\uparrow$ antiferromagnetic   order of spins at uranium atoms in \us\ along the [001] crystallographic direction \cite{Leciejewicz1967,Baek2010}  (cf. Fig. \ref{fig1}) remains enigmatic. Additionally, very recent experiment on this compound \cite{Jeffries2016} has suggested a pressure-induced onset of the ferromagnetic  phase which calls for a theoretical clarification.

In the present work, motivated by these findings, we develop a low-energy microscopic model rationalizing the unusual antiferromagnetism \cite{Leciejewicz1967,Baek2010}  and the pressure-induced transition to ferromagnetism \cite{Jeffries2016} in \us. Relying on the obtained results we also address properties of magnetic states observed in other U-based compounds: isostructural to \us\ UAs$_2$ and UBi$_2$  \cite{Aoki1999,Aoki2000,Aoki2000b,Eriksson2006}, UAsSe \cite{Henkie1994,Henkie1998,Henkie2001} and URh$_x$Ir$_{1-x}$Ge \cite{Aoki2017}.
Moreover, we provide arguments that ferromagnetism in pressurized \us\ has the same character as  ferromagnetism \cite{Wysokinski2014R,Wysokinski2015R,Abram2016} in the famous spin-triplet superconductor \ug\ \cite{Saxena2000}.
The raised analogy between \ug\ and \us\ opens up  an intriguing possibility that \us\ under pressure may also develop triplet superconductivity in yet unexplored  regime of temperature below 2K \cite{Jeffries2016}.

\section{Model} 
\us\  is an antiferromagnet    with N\'eel temperature, $T_N\simeq200K$ \cite{Guziewicz2004} characterized with moderately renormalized effective masses $m^*/m_0\lesssim 8.33$ \cite{Aoki1999,Aoki2000}. Both, quantum oscillations measurements \cite{Aoki1999,Aoki2000} and {\it ab-initio} calculations \cite{Eriksson2006} suggest predominantly two dimensional character of the Fermi surface with a weak dispersion along $z$-direction (cf. Fig.\ref{fig1}). Moreover, the low-energy band structure of \us\ is determined \cite{Eriksson2006} as mostly composed of $5f$ uranium orbitals hybridized to these of the ligand. Such scenario has been also indicated in photoemission spectroscopy measurements \cite{Guziewicz2004,Durakiewicz2008,Beaux2011,Durakiewicz2013usb,Xie2016}. 

These properties together with the initially localized nature of $f$-states, indicated by a spatial separation of U-atoms larger than $4 \mathring{A}$, advocate the effective description of \us\ within the usual Anderson lattice scenario \cite{Wysokinski2015}. Nonetheless, such a description is clearly insufficient in the light of the observation of the unusual  $\uparrow\downarrow\downarrow\uparrow$ antiferromagnetic order of spins at uranium atoms along the $z$-direction \cite{Leciejewicz1967}, as well as a presence of different core levels corresponding to  crystallographically inequivalent ligand sites Sb(I) and Sb(II) \cite{Beaux2011,Kato2004,Baek2010}  (cf. Fig. \ref{fig1}). Therefore, in the following considerations we  account for more specific details of the electronic structure of \us.

First, we take advantage of  a quasi two dimensional character of \us\   \cite{Aoki1999,Aoki2000,Eriksson2006} and consider only a film of the material (parallel to $xy$-plane) of a minimal thickness allowing for a description of the unusual order along $z$-axis. We choose  
thickness corresponding to the height of the elementary cell, i.e., including three neighboring $xy$-planes of uranium atoms and antimony in between.
Within such a probe of \us\ structure we aim to validate that any three consecutive $xy$-planes of uranium atoms develop  [U($\sigma$)Sb(I)U($\sigma$)Sb(II)U($\bar \sigma$)] sequence  along $z$-axis, where $\sigma\in\!\!\{\uparrow\!,\!\downarrow\}$. Here we marked particular positions of inequivalent ligand sites in between  $\sigma$-polarized layers of U atoms. This conclusion would demonstrate that the polarizations of the neighboring uranium planes with Sb(I) sites in between align ferromagnetically, whereas with Sb(II) sites antiferromagnetically. Thus it would provide a direct rationalization of the emerging magnetism in \us.  
 
We describe layers of uranium atoms  with  Hamiltonians $H_{\alpha\in\{1,2,3\}}^f$, whereas two systems of the conduction electrons corresponding to inequivalent sites of ligand with $H_{\alpha\in\{1,2\}}^c$. Assignment of these partial Hamiltonians to 
particular systems of atoms is pictorially presented over the height of the single unit cell  in  Fig. \ref{fig1}. Due to the weak dispersion along $z$-axis we assume coupling only between the nearest systems of orbitals of uranium and antimony, i.e. describing hybridization.
Precisely, we consider coupling of $H_1^c$ to $H_1^f$ and $H_3^f$ through operator $\Gamma_1$, and  analogically of $H_2^c$  to $H_2^f$ and $H_3^f$ through operator $\Gamma_2$ (cf. Fig. \ref{fig1}). Then, the full Hamiltonian    for an isolated film of \us\ reads 
\begin{equation}
\mathcal{H}= (H^c_1+H^f_1+\Gamma_1) + H^f_3+ (H^c_2+H^f_2+\Gamma_2). 
\label{H}
\end{equation}
We propose $\mathcal{H}$ to be a purely two dimensional model. Therefore, the properties along $z$-direction are solely coded in the indices of operators in \eqref{H} referring to the vertical arrangement of considered systems of atoms in accordance with   Fig. \ref{fig1}. 
We emphasize with brackets in Eq. \eqref{H} emerging  picture of two  
semi-independent subsystems coupled only through sharing a single layer of uranium atoms ($H^f_3$). Hereafter, we  use a convention that $H^f_3$ and all operators with index $\alpha=1$ determine the subsystem 1 and the same $H^f_3$ and operators with index $\alpha=2$  constitute the subsystem 2 (cf. Fig. \ref{fig1}).

First we formulate partial Hamiltonians for electrons at uranium layers. These, according to the {\it ab initio} calculations  \cite{Eriksson2006}, provide contribution to the low-energy band structure predominantly of $5f$ orbital character. 
All $f$-states, due to the large separation between uranium atoms, initially occupy  the same atomic energy level $\epsilon_f$. For the reason that  any two neighboring layers of uranium atoms form a bipartite square lattice in $xy$-coordinates, we may propose
\begin{equation}
H_{\alpha\in\{1,2,3\}}^f=\epsilon_{f} \sum_{\g i\in \beta ,\sigma} n^f_{\alpha;\g i\sigma}+U  \sum_{\g i\in \beta}   n^f_{\alpha;\g i\uparrow}   n^f_{\alpha;\g i\downarrow},
\end{equation}
where lattice summation runs over sublattice, $\beta={\rm A}$ for $\alpha=1,2$ and $\beta={\rm B}$ for $\alpha=3$, and $n^f_{\alpha;\g i\sigma}\equiv f^\dagger_{\alpha;\g i\sigma}f^\dagga_{\alpha;\g i\sigma}$  is the number operator of $f$-electrons in $\alpha$-layer of uranium atoms. 
We account for many-body onsite interaction with amplitude $U$ to account for the strongly correlated nature of $f$-states. On the other hand,  to underline the prevailing importance of this interaction we neglect orbital degeneracy of $5f$-shell and thus many-body effects such as the Hund's coupling.  

In turn, conduction electrons, described with  $H_{\alpha\in\{1,2\}}^c$, form bands. For simplicity, we shall assume existence  of only two bands in the vicinity of the Fermi level, each generated either from orbitals at Sb(I) ($H_1^c$) or Sb(II) ($H_2^c$) sites. Relying on experimental observations \cite{Beaux2011} we account for the  shift to the higher energy of the center of Sb(I)-band with respect to this of Sb(II)-band.
On the other hand, we shall disregard other possible differences between bands, e.g, in their shape which should play a minor role due to the strong mixing between $f$ and conduction states. For simplicity we generate both bands with the nearest neighbor hopping with the amplitude $t$ on a square lattice,
\begin{equation}
 H_{\alpha\in\{1,2\}}^c= -t \sum_{\langle\g i \g j\rangle \sigma}  \! c_{\alpha;\g i\sigma}^\dagger  c^\dagga_{\alpha;\g j\sigma}+ \delta_{\alpha,1} \Delta\sum_{\g i \sigma}n^c_{\alpha;\g i\sigma},
\label{delta}
\end{equation}
where $n^c_{\g i\sigma}\equiv c^\dagger_{\g i\sigma}c^\dagga_{\g i\sigma}$  is the number operator for conduction ($c$-) electrons, and $\Delta>0$ is a relative shift between conduction bands.

The remaining, yet unspecified, ingredients of the full Hamiltonian $\mathcal{H}$ \eqref{H}, are the hybridization operators $\Gamma_{\alpha\in\{1,2\}}$.
Due to inequivalent crystallographic positions of Sb(I) and Sb(II) sites 
in principle hybridization functions may differ with a shape or an amplitude. 
 However, for the present purposes it is enough  to assume both operators $\Gamma_{\alpha\in\{1,2\}}$ to describe identical onsite hybridization
\begin{equation}
\begin{split}
\Gamma_{\alpha\in\{1,2\}}\!= \!V \Big[\! \sum_{\g i\in A,\sigma}\!  f_{\alpha;\g i\sigma}^\dagger
  c^\dagga_{\alpha;\g i\sigma}\! +\!\!\sum_{\g i\in B,\sigma}\! f_{3;\g i\sigma}^\dagger
  c^\dagga_{\alpha;\g i\sigma}\Big] + {\rm H.c.}   
\end{split}
\end{equation}

Having determined form of all terms in the model \eqref{H}, we may conclude that the proposed earlier two semi-independent subsystems are two Anderson lattices sharing one of the sublattices  of $f$-orbitals (here B).
Until now the only apparent difference between both subsystems is the relative shift ($\Delta$)  between the centers of the conduction bands, supported by experimental observations of different 
core levels corresponding to  crystallographically inequivalent ligand sites Sb(I) and Sb(II) \cite{Beaux2011,Kato2004,Baek2010}  (cf. Fig. \ref{fig1}).
However, in a consequence of a common chemical potential in both subsystems, a total number of electrons in each of them ($n^t_{\alpha\in\{1,2\}}$) is different in favor of the subsystem 2 ($n^t_{2}>n^t_{1}$). Namely, although $f$-electron number in each of the subsystems is the same, the conduction band associated with the subsystem 2  due to the lower position of its bottom accommodates more electrons.
We shall demonstrate that the presence of mentioned differences between subsystems (shift between conduction bands and filling difference) is a sufficient condition to account for the nonstandard magnetic properties of \us.

Coupling between subsystems is introduced solely via sharing one of the $f$-electron sublattice (cf. Fig. \ref{fig1}). Such circumstance indicates that both subsystems may be treated independently as long as $f$ electron number per spin at sublattice B is the same for both of them. In result we shall consider magnetic properties of each of the subsystems separately with the following Hamiltonians
\begin{equation} 
\begin{split}
& H^*_{\alpha\in\{1,2\}}=H_\alpha^c+H_\alpha^f + \Gamma_\alpha+H_3^f\\
&=\!   -t \!  \sum_{\langle\g i \g j\rangle,\sigma} \!    c_{\g i\sigma}^\dagger  c^\dagga_{\g j\sigma}+\delta_{\alpha1} \Delta\! \sum_{\g i \sigma}\! n^c_{\g i\sigma}   + V  \! \sum_{\g i\sigma}\! ( f_{\g i\sigma}^\dagger
  c^\dagga_{\g i\sigma}\!+\!{\rm H.c.})\\
& \ \ \ + \epsilon_{f} \sum_{\g i ,\sigma} n^f_{\g i\sigma}+
U  \sum_{\g i}   n^f_{\g i\uparrow}   n^f_{\g i\downarrow}-\mu_\alpha \sum_{\g i ,\sigma} (n^f_{\g i\sigma}+n^c_{\g i\sigma}).
\label{RH}
\end{split}
\end{equation}
Coupling between subsystems is simply reestablished by a matching of the average number of $f$-electrons per spin at the sublattice B resulting from consideration of $H^*_1$ and $H^*_2$ separately. 
\

\section{Applicability of the model to $\rm\mbox{\us}$}
\subsection{Renormalized mean-field solution}
We approach Hamiltonian \eqref{RH} in a mean-field manner accounting for antiferro- and ferro-magnetic solutions.
Antiferromagnetic state is characterized with a spatial modulation of the number of electrons per spin $n_{\g i \sigma}\!=\! \frac{1}{2}(n^t\!+\!	\sigma m_{AF} {\rm e}^{i \g Q \g R_{\g i}})$ with  $\sigma\!\equiv\!\{+,-\}$ for spin $\{\uparrow,\downarrow\}$, where ${\g Q\!=\! (\pi,\pi)}$ is the usual ordering vector and $m_{AF}$ staggered magnetization.
In turn, the ferromagnetic state is characterized with  $n_{\g i \sigma}\!=\! \frac{1}{2}(n^t\!+\!\sigma m_{FM})$ where $m_{FM}$ is uniform spin polarization.
For actual calculations (at zero temperature) we use the Gutzwiller approximation combined with a self-consistent optimization of the Slater determinant  \cite{Abram2013,Zegrodnik2013,Zegrodnik2014,Wysokinski2014,Abram2017}, equivalent to slave-boson technique \cite{Gebhard2007}.  
Details of the method for the consideration of ferromagnetism are presented in  Ref. \onlinecite{Wysokinski2014R}.
Therefore, here we present only variational  antiferromagnetic Hamiltonian defined in the reduced  Brillouin zone (RBZ)  
\widetext
\begin{equation}
H^{AF} =\sum_\sigma\int_{\g k \in RBZ}\!\! d\g k\ 
\Psi^\dagger
 \begin{pmatrix}
  \epsilon_{\g k}-\mu & 0  &\frac{q_{\sigma}+q_{\bar\sigma}}{2}V & \frac{q_{\sigma}-q_{\bar\sigma}}{2}V\\
 0 & \epsilon_{\g k+\g Q}-\mu & \frac{q_{\sigma}-q_{\bar\sigma}}{2}V&\frac{q_{\sigma}+q_{\bar\sigma}}{2}V\\
  \frac{q_{\sigma}+q_{\bar\sigma}}{2}V & \frac{q_{\sigma}-q_{\bar\sigma}}{2}V & \epsilon_f-\mu+ \frac{\lambda_{\sigma}+\lambda_{\bar\sigma}}{2}  & \frac{\lambda_{\sigma}-\lambda_{\bar\sigma}}{2} \\
  \frac{q_{\sigma}-q_{\bar\sigma}}{2}V&\frac{q_{\sigma}+q_{\bar\sigma}}{2}V & \frac{\lambda_{\sigma}-\lambda_{\bar\sigma}}{2}  & \epsilon_f-\mu +\frac{\lambda_{\sigma}+\lambda_{\bar\sigma}}{2} \\
 \end{pmatrix} \Psi,
\end{equation}
\endwidetext \noindent
where $\Psi^\dagger=(c^\dagger_{\g k \sigma},c^\dagger_{\g k+\g Q \sigma},f^\dagger_{\g k \sigma},f^\dagger_{\g k+\g Q \sigma})$, $\epsilon_{\g k}$ is the tight binding spectrum of a conduction band, $q_\sigma$ is the usual Gutzwiller narrowing factor \cite{Wysokinski2014,Wysokinski2014R}, and
\begin{equation}
\lambda_\sigma= \frac{\partial \langle H^*\rangle_G}{\partial \langle n_{\sigma}\rangle_0}.
\end{equation}
 Here $\langle ...\rangle_0$ denotes an expectation value with the Slater determinant and $\langle ...\rangle_G$ with the Gutzwiller wave function under the Gutzwiller approximation.
 
 In calculations, we choose hopping of conduction electrons as the energy scale, i.e., $t=1$. If it is not stated otherwise remaining microscopic parameters are set to $U=10$, $V=-0.7$, $\epsilon_f=-2.5$. Large $f$-$f$ interaction and $f$-level well below the center of the conduction band are indirectly suggested by the effective quasiparticle mass renormalization in \us\ \cite{Aoki1999,Aoki2000}. On the other hand sizable hybridization has been suggested by the photoemission spectroscopy \cite{Guziewicz2004}.
 The other important parameter, a total number of electrons in both subsystems $n^t_{1}+n^t_{2}$ is an arbitrary chosen fitting parameter. The convenient fitting procedure can be established by setting filling in only one of the subsystems (e.g. $n^{t}_2$) and deriving $n^t_1+n^t_2$ afterwards from the matching condition between subsystems.   

 \subsection{Ambient pressure groundstate}
 
It is a well-established result that if the total number of electrons   is close to the half-filling, the Anderson lattice model supports antiferromagnetism  \cite{Doradzinski1997,Doradzinski1998,Kubo2015}. In turn,  if the total number of electrons is away from the half-filling it favors ferromagnetism instead \cite{Doradzinski1997,Doradzinski1998,Wysokinski2014R,Kubo2015, Bonca1,Bonca2,Bonca3}. Therefore, for the subsystem 2 we set number of electrons per site close to the half-filling, $n^t_{2}=1.9$ where indeed antiferromagnetic state is stable against  ferromagnetism and paramagnetism and is characterized with $f$-electron number per site $n^f_2=0.980$ and staggered $f$-electron magnetization  $m_{f;AF}=0.973$.
Now, parameters characterizing the subsystem 1 ($\Delta>0$ and $n^t_{1}<n^t_{2}$) are determined by reintroducing its coupling to the subsystem 2. 
Namely, we look for $\Delta$ and $n^t_{1}$ such that the ferromagnetic solution would be characterized with the $f$-electron number $n^f_1=0.980$ and uniform $f$ electron magnetization $m_{f;FM}=0.973$.
This procedure  yields $n^t_{1}\simeq 1.7$ and $\Delta\simeq0.65$. In this case however, ferromagnetism has the lowest energy  among considered states. 
Such result demonstrates that the alternation in positions of centers of conduction bands, associated with Sb(I) and Sb(II) ligand sites is sufficient feature of this material that can be responsible for the particular nonstandard antiferromagnetic order  observed in \us\ at ambient pressure \cite{Leciejewicz1967}.

 \begin{figure}[t]
  \begin{center}   
     \includegraphics[clip,trim=0cm 0cm 0cm 0.01cm ,width= 0.5  \textwidth]{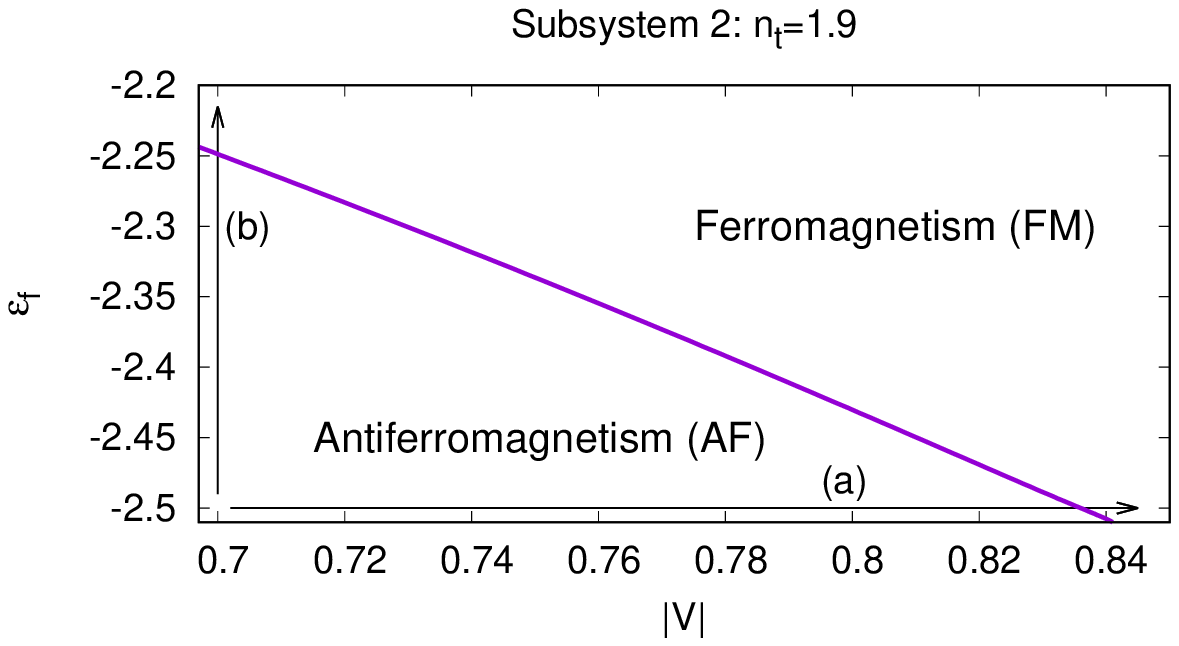}
  \includegraphics[width= 0.23 \textwidth]{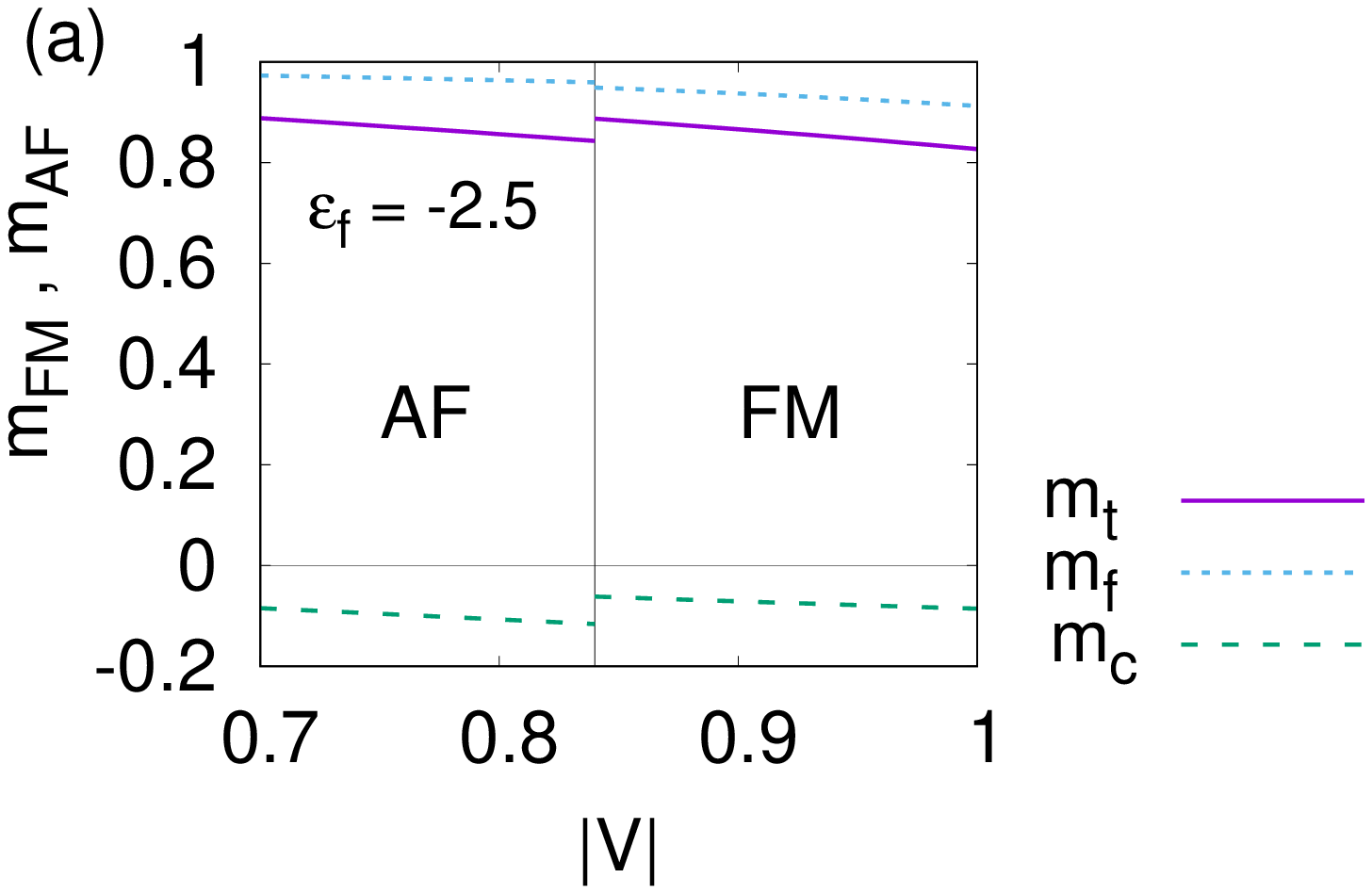}
     \includegraphics[width= 0.23  \textwidth]{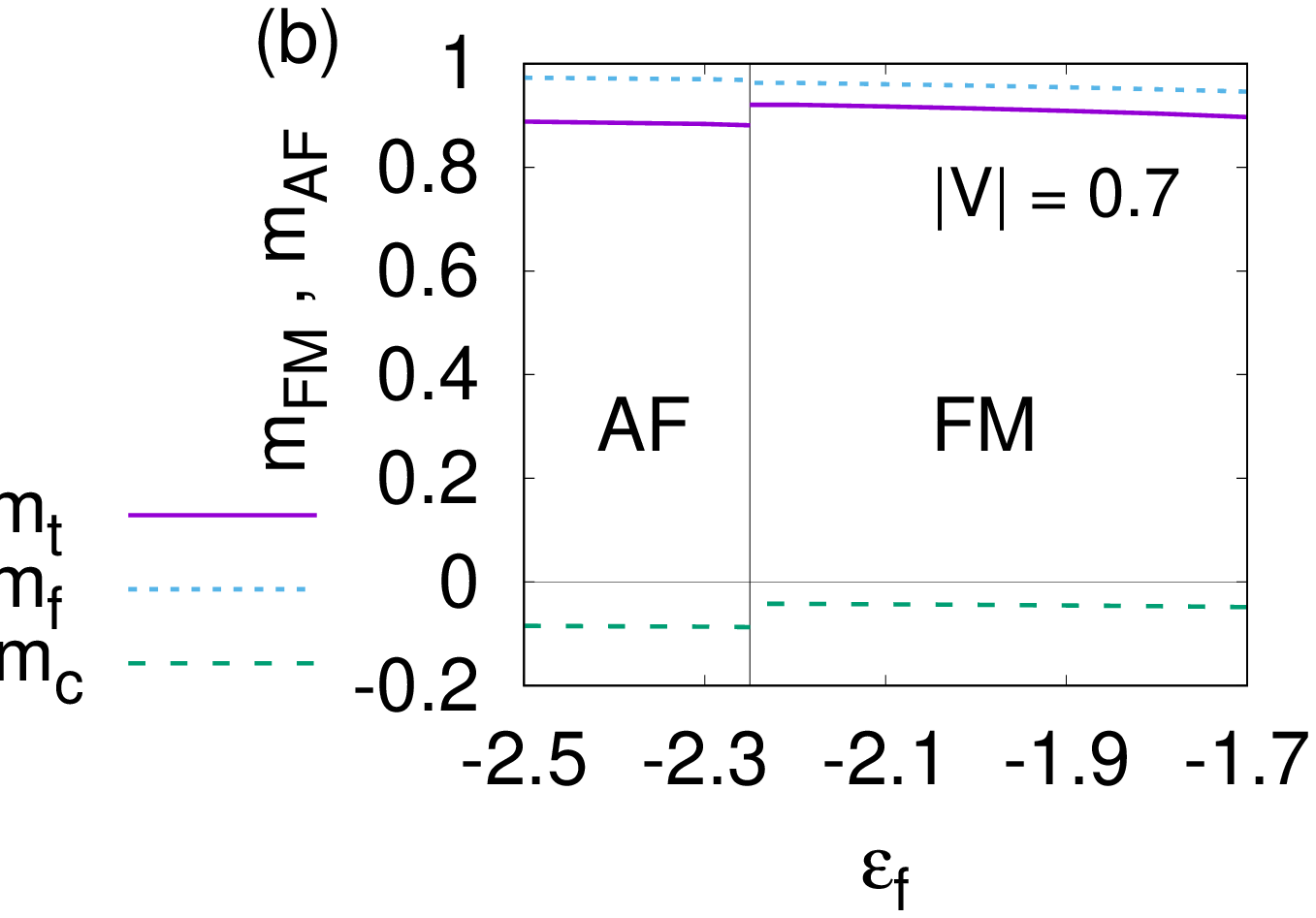}
  \end{center} \vspace{-0.3cm}
 \caption{Phase diagram on $\epsilon_f-|V|$ plane resulting from Hamiltonian \eqref{RH}   for electron filling $n^t=1.9$ and $\Delta=0$. Total number of electrons as well as parameter's values $\epsilon_f=-2.5$ and $|V|=0.7$ (left, bottom corner of diagram) refer to the ambient pressure state of the subsystem 2. In panels (a) and (b) there are plotted orbitally resolved magnetization curves ($m_c$ for  conduction electrons, $m_f$ for $f$-electrons and $m_t=m_c+m_f$) across the phase transition along selected, and marked with arrows  at the phase diagram, directions.} 
\label{fig2} 
\end{figure}

 \subsection{Magnetic groundstate switching under pressure}
Now we shall discuss the evolution of the predicted groundstate of \us\ driven by the change of microscopic parameters associated with the increase of pressure in the considered system. 
In the Anderson lattice model, impact of pressure is mostly associated with the decreasing ratio $U/V$ or $\epsilon_f/V$ \cite{Wysokinski2014R,Wysokinski2015R,Kubo2013,Lacroix1999}. In principle a relation of these parameters to the conduction band-width should be also important, and therefore  fixed ratio $V/t$ during the evolution with pressure is sometimes considered \cite{Lacroix1999}.  Here however, for simplicity, the effect of pressure is modeled by the increase of the hybridization amplitude $|V|$  and the increase of the atomic $f$-electron level $\epsilon_f$, given that the remaining microscopic parameters do not change. 
In Fig. \ref{fig2} we present magnetic phase diagram of the Anderson lattice on the $\epsilon_f$-$|V|$ plane for fixed total number of electron $n^t=1.9$  corresponding to the initial filling of the subsystem 2.
Increase of either $|V|$ or $\epsilon_f$ inevitably drives the transition from antiferromagnetism to ferromagnetism. On the other hand, the Anderson lattice for $n^t=1.7$  corresponding to the initial filling of the subsystem 1, in the range of parameters presented in Fig. \ref{fig2} orders ferromagnetically. 
This result explains the transition to the ferromagnetic state observed in USb$_2$ \cite{Jeffries2016} under applied pressure.

Here we have assumed that number of electrons is conserved
separately in each of the subsystems. 
Nevertheless, we have also checked that if a difference between fillings is decreasing along any hypothetical line associated here with applied pressure and $\Delta>0$,  reached conclusions are the same.

In order to characterize, at least partially, obtained magnetic groundstate switching from antiferromagnetic to ferromagnetic one in Figs. \ref{fig2}(a-b) we plot total ($m_t$)  as well as orbitally-resolved  magnetizations ($m_c$ of conduction electrons, $m_f$ of $f$-electrons) across such transition for selected directions marked at phase diagram with arrows. It is quite intriguing that although magnetic groundstate switches, absolute value of spin-polarization at single site changes only slightly.
We believe that this feature could be verified experimentally. In fact similar behavior, i.e. comparable uniform and staggered magnetizations across phase transition between different  magnetic groundstates, has been recently observed in LaCrGe$_3$ \cite{Taufour2016,Taufour2017}.
 
Further increase of hybridization  or raise of $f$-level in proposed model should drive the transition from  ferromagnetic to  paramagnetic phase \cite{Wysokinski2014R,Kubo2013} as it is also observed in \us\ under pressure \cite{Jeffries2016}.
However, within the considered here disentangled \eqref{RH} rather than full model \eqref{H}, for instance the order of such transition is not resolvable in a sensible manner. 
Nonetheless, as a consequence of predicted in a present work itinerant character of ferromagnetism in \us\ we may safely predict the existence of tricritical wings in this material, semi-universally observed in $d$- and $f$-electron \cite{Uhlarz2004,Taufour2010,Taufour2017,Brando2016}  metallic ferromagnets. These are theoretically supported in the Anderson lattice already at the mean-field level given that the transition to paramagnetism is inherently of the first order \cite{Wysokinski2015R,Abram2016}. On the other hand, if the transition at the mean-field level would be of the second order, than under the theory of Belitz, Kirkpatrick and Vojta  quantum criticality would be avoided \cite{Belitz1997} and tricritical wings would nevertheless appear \cite{Belitz2005}.

\section{Correspondence of the model to other U-based materials}
In the present work we have constructed a microscopic model for magnetism observed in \us. 
However, there are also other materials that the properties of the proposed model can refer to. These are isostructural to \us\ UAs$_2$ and UBi$_2$, UAsSe ,  URh$_x$Ir$_{1-x}$Ge and famous spin-triplet superconductor UGe$_2$. In this section we shall comment on the magnetic properties of these materials in a context of obtained results.   
\subsection{\ua\ and \ub\ }
Isostructural uranium dipnictides UX$_2$ (X=Sb,As,Bi) share many similarities, starting from (i) antiferromagnetic order at ambient pressure, (ii) quasi-two dimensional, cylindrical Fermi surfaces along [001] direction \cite{Aoki1999,Aoki2000,Aoki2000b}, (iii)  strong evidences for the hybridization of $5f$ with conduction  electrons \cite{Guziewicz2004,Eriksson2006}, as well as (iv) uranium atom spacing beyond so-called Hill limit. We have found in the literature clear evidences for the different core levels of inequivalent ligand sites only for \us\ \cite{Beaux2011,Kato2004,Baek2010}. However, in a following discussion we shall assume that the same scenario applies also to the other members of the series, and thus proposed in a present work model is relevant for their description as well. 

\begin{table}[b]
 \begin{tabular}{ |c|c|c|c|c|c| } 
 \hline
       &$T_N$ (K) & $\mu_{\rm ord}$ ($\mu_B$/U)& [001]-seq. &$m^*/m_0$  \\
\hline  \ua\ & 273 & 1.61 & ($\uparrow\downarrow\downarrow\uparrow$)&0.33-4.50\\ 
\hline  \us\ & 200 & 1.88 & ($\uparrow\downarrow\downarrow\uparrow$)&1.84-8.33\\  
\hline  \ub\ & 183 & 2.1  & ($\uparrow\downarrow\uparrow\downarrow$)&4.40-9.20\\
\hline 
\end{tabular}
\caption{Differences among \ux\ series in Neel temperature ($T_N$), ordered moment ($\mu_{\rm ord}$), sequence of polarization of uranium layers along [001] direction and effective mass renormalization ($m^*/m_0$) as seen by de Haas-van Alphen oscillations.}
\label{t1}
\end{table}

In Table \ref{t1} we have gathered selected  
properties of members of \ux\ series organized with decreasing $T_N$.
Apparently \ub, although it has lowest N\'eel temperature, is a better antiferromagnet than \us, and is characterized with the larger moment at uranium atoms forming usual  sequence of spin polarizations $\uparrow\downarrow\uparrow\downarrow$ along $z$-axis. On the other hand, based on the same characteristics \ua\ seems to be the worst antiferromagnet among \ux\ family. 

Increasing effective masses when going from \ua\ through \us\ to \ub\  can point to a possible decrease of the hybridization amplitude \cite{Wysokinski2015}.  Such scenario can be indeed realized for UX$_2$ family because $5f$ electrons hybridize mostly to conduction electrons derived from $6p$ orbitals in \ub, $5p$ in \us, and $4p$ in \ua\ \cite{Eriksson2006}. Due to the widest  bandwidth originating from highly delocalized $6p$ orbitals in \ub\ the relative ratio of hybridization to conduction electron bandwidth can be expected to be small. Consequently, by virtue of the same argument \ua\ can be characterized with a relatively large effective hybridization in units of conduction bandwidth.
Therefore, one can expect that small hybridization,  which usually favors antiferromagnetic orderings even quite away from half filling in Anderson lattice  \cite{Doradzinski1997,Doradzinski1998}, leads to antiferromagnetic order in both semi-independent subsystems in \ub\ providing observed standard $\uparrow\downarrow\uparrow\downarrow$ ordering in contrast to a situation in \us\ and \ua.

\subsection{UAsSe}
UAsSe is a ferromagnet at ambient pressure \cite{Henkie2001}. By comparing crystal structures of \us\ with UAsSe, the arsenic and selenium atoms in the latter compound play the role of Sb(I) and Sb(II) atoms respectively in \us. In turn, photoemission studies \cite{Guziewicz2006} suggest that the delocalization of 5f electrons is direction-dependent, i.e., $f$-states strongly contribute to the band dispersion within $x-y$ plane while have localized nature along $z$-axis. 
These properties of UAsSe support a formation of two semi-independent subsystems, weakly dispersive along $z$-axis, and thus coupled mostly through sharing single layer of uranium atoms in a same manner as we propose in a present work for \us.

The emergence of uniform ferromagnetic polarization in UAsSe instead of non-standard antiferromagnetic ordering in \us\ can be attributed to many potential reasons, though here we address only one.  The emergent conduction electron bands derived from arsenic and selenium atoms in UAsSe can even more significantly differ in hybridization functions and relative shift $\Delta$ than it is the case for these derived from Sb(I) and Sb(II) atoms in \us. Such circumstance may lead to the situation in which one of the subsystems is characterized with the filling sizably lower and the other sizably higher than half-filling. This may follow that ferromagnetism can be more stable against antiferromagnetism that is usually favored in the  vicinity of half-filling.  It must be noted however that a present theoretical approach is not well-suited to address UAsSe for its complete set of properties.
Namely, the Kondo-signatures \cite{Henkie1994,Henkie1998} cannot be addressed within renormalized mean-field theory which treats quantum fluctuations in an oversimplified manner to account properly for this class of many-body effects.

\subsection{URh$_x$Ir$_{1-x}$Ge}
URhGe is an itinerant ferromagnet that reflects unusual coexistence of uniform polarization with a spin-triplet superconductivity \cite{Aoki2001,Aoki2012}. There are evidences for a  predominant mechanism for a delocalization of 5f electrons based on a hybridization to other orbitals in this material, presumably $4d$ of ruthenium \cite{Fujimori2014}. Although, 
the distance between neighboring uranium atoms is at the border of the so-called Hill-limit to 
unambiguously neglect direct $f$-$f$ electron hopping, in a following we shall assume that effectively ferromagnetism in URhGe can be described by the Anderson lattice model (cf. Eq. \eqref{RH}). 
In that manner the results of Fig. \ref{fig2} derived for such a model may apply to the case of this material. 

In a recent experiments there has been thoroughly examined evolution with doping of the magnetic groundstate of URhGe when rhodium is exchanged for iridium \cite{Aoki2017}. 
At 56\% content of Ir (URh$_x$Ir$_{1-x}$Ge, $x=0.56$)  the compound undergoes transition to antiferromagnetic state with a very similar staggered and uniform magnetizations at both sides of a transition \cite{Aoki2017}. Doping with iridium may gradually change the character of conduction states from 4$d$ to 5$d$, and therefore to increase of their bandwidth. In result effective hybridization in a units of bandwidth should decrease, what according to results presented in Fig. \ref{fig2}(a) can rationalize  magnetic ground state switch observed in URh$_x$Ir$_{1-x}$Ge \cite{Aoki2017}.
 
\subsection{\ug}
In the present work, the character of ferromagnetism in pressurized \us\ is proposed to originate from the competition between $c-f$ hybridization and $f-f$ interaction energy scales within the framework 
of Anderson lattice model. Previously  we have proposed \cite{Wysokinski2014R,Wysokinski2015R,Abram2016} the same mechanism for ferromagnetism in  the spin-triplet superconductor \ug\ \cite{Saxena2000}.
According to our present choice of parameters, high pressure ferromagnetic phase in \us\ has a character similar to the strongly polarized phase (FM1) of  \ug\  \cite{Wysokinski2014R}. 

There exists  a number of observations suggesting that ferromagnetism and superconductivity in \ug\  are strongly intertwined \cite{Aoki2012}. It may follow  that, the  onset of  superconductivity is related to the particular mechanism driving the ferromagnetic state itself.
In result, advocated scenario of the same mechanism leading to appearance of ferromagnetism in pressurized \us\ and in \ug, may hint that the former compound also develops triplet superconductivity in yet experimentally unexplored regime \cite{Jeffries2016} of temperature below 2K. 

\
     
\section{Summary} 
In the present work we have developed low-energy model which at the mean-field level explains magnetic properties of \us:  the unusual $\uparrow\downarrow\downarrow\uparrow$ sequence of polarizations at uranium atoms  along [001] crystallographic direction   \cite{Leciejewicz1967,Baek2010} and the pressure induced transition to the ferromagnetic state \cite{Jeffries2016}. 
Relying on the obtained results for the proposed model we have addressed some of the properties of magnetic states also in other U-based compounds: UAs$_2$ and UBi$_2$  \cite{Aoki1999,Aoki2000,Aoki2000b,Eriksson2006}, UAsSe \cite{Henkie1994,Henkie1998,Henkie2001} and URh$_x$Ir$_{1-x}$Ge \cite{Aoki2017}.
Moreover, our modeling suggests a strong analogy between ferromagnetic phases of \us\ and spin-triplet superconductor \ug\ \cite{Wysokinski2014R, Wysokinski2015R}. Finally, we note that the present work can guide future studies toward explanation of other intriguing observations available for \us\ such as a presence of a kink in $f$-electron dispersion \cite{Durakiewicz2008,Durakiewicz2013usb} and $T$-linear scattering rate in the high-pressure paramagnetic phase \cite{Jeffries2016}.

\section*{Acknowledgments}
The stimulating discussions with 
M. Abram, A. Amaricci, D. Aoki, W. Brzezicki, M. Capone, M. Fabrizio  and G. Knebel are greatly acknowledged.
This work has been supported by the Polish Ministry of Science and Higher Education under the ``Mobility Plus'' program, Agreement No. 1265/MOB/IV/2015/0, as well as by the Foundation for Polish Science through  ``START'' fellowship as well as through the IRA Programme co-financed by EU under the European Regional Development Fund.

 \end{document}